# Introduction to Longitudinal Beam Dynamics

*B.J. Holzer*
CERN, Geneva, Switzerland

**Abstract**
This chapter gives an overview of the longitudinal dynamics of the particles in an accelerator and, closely related to that, the issue of synchronization between the particles and the accelerating field. Beginning with the trivial case of electrostatic accelerators, the synchronization condition is explained for a number of driven accelerators like Alvarez linacs, cyclotrons and finally synchrotrons and storage rings, where it plays a crucial role. In the case of the latter, the principle of phase focusing is motivated qualitatively as well as on a mathematically more correct level and the problem of operation below and above the transition energy is discussed. Throughout, the main emphasis is more on physical understanding rather than on a mathematically rigorous treatment.

## 1  Introduction

The longitudinal movement of the particles in a storage ring – and, closely related to this, the acceleration of the beam – is strongly related to the problem of synchronization between the particles and the accelerating system. This synchronization might be established via the basic hardware and design of the machine (like in a Wideroe structure), via the orbit (in a cyclotron), or in a more sophisticated way it can be a fundamental feature of the ring, which leads in the end to the name 'synchrotron'. We will treat these different aspects in more detail. But, before we do, we would like to start on a very fundamental ground and at the same time go back a bit in history.

## 2  Electrostatic machines

The most prominent example of these machines, besides the Cockcroft–Walton generator, is the Van de Graaff generator. A sketch of the principle is shown in Fig. 1.

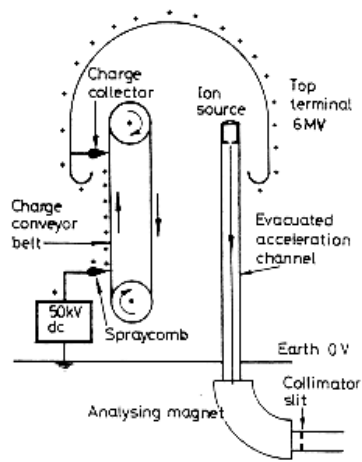

**Fig. 1:** Technical principle of a Van de Graaff accelerator

Using a moderate d.c. high voltage, charges are sprayed on a moving belt or chain with isolated links and transported to a kind of Faraday screen, where the charges accumulate, leading to considerable high voltages. Basically, it is the transformation of mechanical energy into electrostatic energy via the movement of the belt that defines the physical principle behind the generator. The high-voltage part is usually connected to a particle source (protons or heavy ions) and the potential difference between the high-voltage end and the ground potential will define the particle energy. Voltages in the range of some megavolts can be achieved, where usually a discharge-suppressing gas is needed to achieve the highest possible terminal voltages (up to ≈ 30 MV). By their design concept, these machines deliver an excellent energy resolution, as basically each and every particle sees the same accelerating potential (which is no longer the case as soon as we have to consider a.c. accelerating structures).

An example of such a machine is shown in Fig. 2. This 'tandem' Van de Graaff accelerator uses a stripper foil in the middle of the structure, thus, in the first half of the structure, accelerating negative ions that are produced in a Cs-loaded source. After stripping the electrons, the same voltage is applied in the second part of the structure to gain another step in energy. However, after leaving the accelerator at the down end, the particle energy is still limited by the high voltage that can be created, and that is finally always limited due to discharge effects of the electric field.

The kinetic energy of the particle beam is given by the integration over the electric field $E$, in direction $z$ of the particle motion and measured as usual in units of electronvolts (eV):

$$dW = eE_z ds \quad \rightarrow \quad W = e \int E_z ds = [\text{eV}]$$

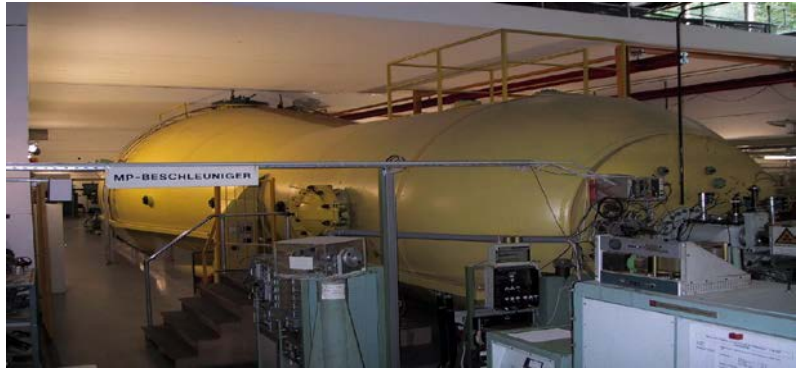

**Fig. 2**: Tandem Van de Graaff accelerator of the MPI Institute, Heidelberg. The ion source, delivering negative ions, is connected to the accelerator on the right-hand side of the picture. The acceleration structure is housed in a vessel filled with discharge suppressing gas (SF6). At the far end, the dipole spectrometer magnet is visible to separate different ion species.

It should be emphasized here that a special synchronization between accelerating voltage and particle beam is not needed as a d.c. voltage is used for acceleration.

## 3  The first radio-frequency accelerator: Wideroe linac

The basic limitation given by the maximum achievable voltage in electrostatic accelerators can be overcome by applying a.c. voltages. However, a more complicated design is needed to prevent the particles from being decelerated during the negative half-wave of the radio-frequency (RF) system. In 1928 Wideroe presented for the first time the layout of such a machine (and built it). Figure 3 shows the principle.

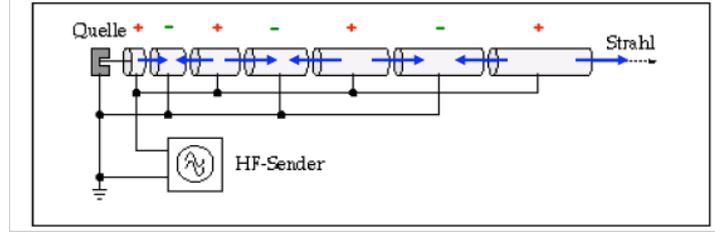

**Fig. 3:** Schematic layout of an RF linear accelerator (drift tube linac) as proposed by Wideroe in 1928

The arrows show the direction of the electric field for a given moment in time (i.e., the positive half-wave of the a.c. voltage) and the corresponding polarity applied to the electrodes indicated by the '+/−' sign in the figure. In principle, arbitrarily high beam energies can be achieved by applying the same voltage to a given number of electrodes, provided that the particle beam is shielded from the electric field during the negative RF half-wave. Accordingly the electrodes are designed as drift tubes ('drift tube linac') whose length is adopted according to the particle velocity and RF period. Schematically the problem is shown in Fig. 4.

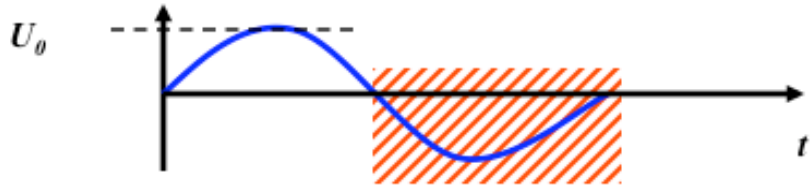

**Fig 4**: During the negative half-wave of the RF system the particles have to be shielded to avoid deceleration

The red area corresponds to the time during which the particle has to be shielded from the decelerating field direction and is defined by the RF period: $t_{\text{shield}} = \tau_{RF}/2$. Accordingly the drift tube length has to be

$$l_i = v_i \cdot \frac{\tau_{rf}}{2}$$

If the kinetic energy (in the classical regime) is given by

$$E_i = \frac{1}{2} m v^2$$

we obtain an equation for the drift tube length that, for a given accelerating voltage $U_0$ and charge $q$, depends on the RF frequency $v_{RF}$ and the accelerating step $i$:

$$l_i = \frac{1}{v_{rf}} \cdot \sqrt{\frac{i \cdot q \cdot U_0 \cdot \sin\psi_s}{2m}} \tag{1}$$

The parameter $\psi_s$ describes the so-called synchronous phase and can be chosen to be 0° in this case to obtain highest acceleration performance.

For completeness, it should be mentioned that in 1946 Alvarez improved the concept by embedding the structure inside a vessel to reduce RF losses. The drift tube linac is typically used as an accelerator for proton and heavy-ion beams, and beam energies of the order of some tens of Mega electronvolt (MeV) are obtained. At the time of writing, a new proton linac is installed at CERN to improve the beam performance for the LHC beams and – as you would expect – again an Alvarez drift tube linac is used to gain in three stages a kinetic energy of the protons of 150 MeV.

One of the best-known examples of such an accelerator is running at GSI in Darmstadt, used as a universal tool for the acceleration of (almost) any heavy-ion species. In Fig. 5 the inner structure is shown, with the drift tubes and the surrounding vessel.

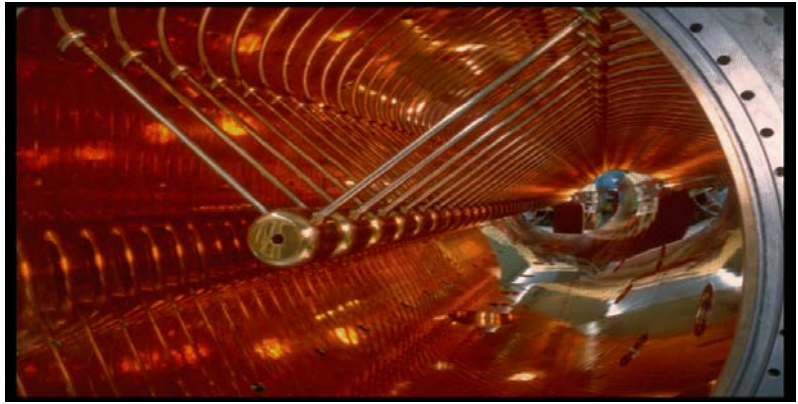

**Fig. 5:** 'Unilac' of GSI

Unlike the d.c. accelerators, synchronization has to be obtained between the particles and the accelerating RF field, which – as shown above – is represented by the drift tube length, and so in a certain sense built into the hardware of the system.

## 4   Cyclotrons

In principle, the drift tube linac concept can be applied to relatively high energies (doing the calculations in Eq. (1), relativistically correct even beyond the classical regime). The length of the drift tubes, however, makes the design very inefficient for high energies, and more sophisticated ideas are used to keep the structure compact and within reasonable financial and technical limits. The idea that is considered as a next natural step is the application of a magnetic dipole field to bend the particle orbit into a circle or, better (as we will see), into a spiral.

A constant dipole field applied perpendicular to the particle motion will lead to a transverse deflecting Lorentz force that is given by

$$\mathbf{F} = q \cdot (\mathbf{v} \times \mathbf{B}) = q \cdot v \cdot B$$

where we have replaced the cross-product by a simple multiplication.

The condition for a circular orbit is given by the equality of this Lorentz force and the centrifugal force, and leads to a relation that is of major importance for any circular machine:

$$q \cdot v \cdot B = \frac{m \cdot v^2}{R} \quad \rightarrow \quad B \cdot R = p/q \tag{2}$$

The expression $B \cdot R$ on the right-hand side is called the beam rigidity, as it describes the resistance of the particle beam to any external deflecting force and clearly depends on the particle momentum. At the same time it defines – for a given size of the machine ($R$) and for a given maximum magnetic field ($B$) of the dipoles installed – the highest momentum that can be carried by the (circular) accelerator.

The revolution frequency of the particles

$$\omega_z = \frac{v}{R} = \frac{q}{m} \cdot B_z \tag{3}$$

is constant as long as we can consider the classical approach, i.e., as long as the mass of the particles can be assumed to be constant. Accordingly the RF frequency applied for the acceleration can be kept constant and – in the ideal case – set equal to a multiple of the revolution frequency to obtain particle synchronization.

A schematic layout of a cyclotron is shown in Fig. 6. The accelerating RF voltage is applied between the two halves of the pill-box-like structure, leading to a step in energy (or better momentum) twice per revolution.

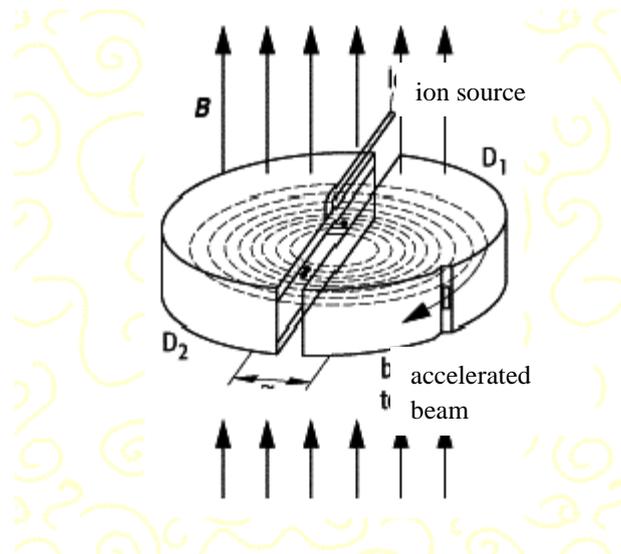

**Fig. 6:** Schematic layout of a cyclotron with the two half-pill-box-like electrodes that contain the spiralling orbit, and the external magnetic field line indicated.

According to Eq. (2), the radius of the particle orbit will increase due to the constant magnetic field, and a spiralling orbit is obtained. The maximum energy achievable in a cyclotron is determined by the strength and geometrical size of the dipole magnet. An example of such a classical cyclotron is shown in Fig. 7. The vacuum chamber of the cookie-like box, the magnet coils and the dipole magnet are clearly visible.

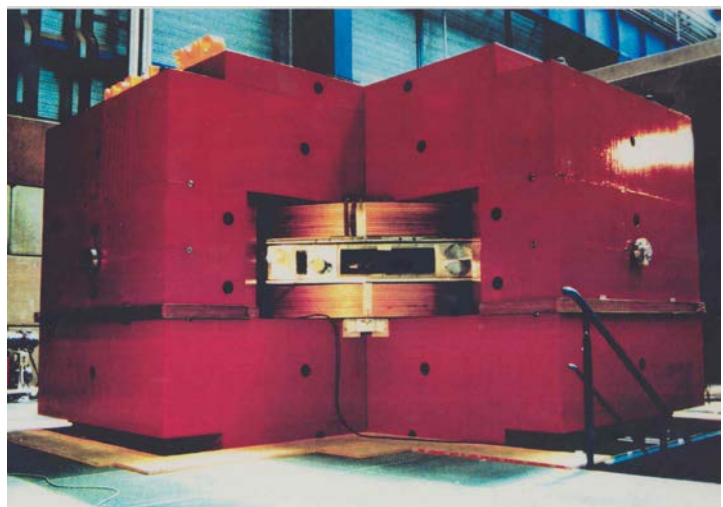

**Fig. 7:** Cyclotron SPIRAL at Ganil

Another limitation, however, should be mentioned. While the revolution frequency of the particles can be considered as constant in the classical treatment of Eq. (3), the situation changes as soon as a considerable mass increase is obtained due to the relativistic particle energies. The correct expression for the revolution frequency $\omega_s$, and thus as well for the RF frequency $\omega_{RF}$, therefore is

$$\omega_s(t) = \omega_{rf}(t) = \frac{q}{\gamma(t) \cdot m_0} \cdot B$$

where $\gamma$ describes the time-dependent relativistic parameter. During the acceleration process the particle frequency slows down, and to keep the particle motion synchronous with the applied RF frequency the external RF system has to be tuned to follow this effect. While higher particle energies can be obtained, the performance of the machine will degrade as a d.c.-like beam acceleration is no longer possible due to the RF cycling that is needed for each acceleration process.

As before, to refer to the problem of synchronization: the synchronous condition between particles and accelerating RF voltage in a cyclotron is defined by the length of the spiralling orbit.

## 5 Radio-frequency resonators and transit time factor

Synchrotrons are ring-like accelerators that use located RF systems for particle acceleration and, owing to their design principle, can achieve up to now the highest particle energies. The RF voltage – much like in the case of a drift tube linac – is usually created in a so-called resonator where a standing RF wave is created, with an electric field vector pointing in the direction of particle motion (Fig. 8).

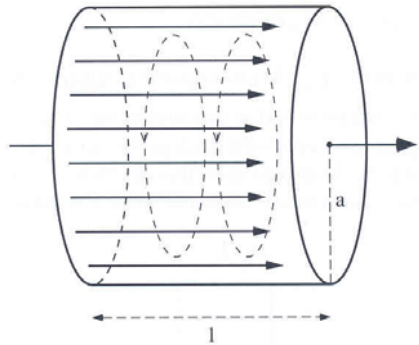

**Fig. 8:** Pill box cavity with longitudinal electric field

Before going into the detail of the particle dynamics in a synchrotron, a problem has to be mentioned that is present in all RF-like structures. The gain in kinetic energy (which is equal in this case to the gain in overall energy) in a d.c. accelerating system is given by

$$dW = dE = eE_z ds$$

with $E_z$ describing the longitudinal component of the electric field, and the overall energy gain $W$ is obtained via a simple integration over the complete path d$s$ of the particle:

$$W = e \int E_z ds = eV$$

The situation changes quite a bit if RF voltage is applied. The electric field is changing as a function of the RF frequency used:

$$E_z = E_0 \cos\omega t = \frac{V}{g}\cos\omega t$$

and the total energy gain obtained is given by

$$\Delta W = \frac{eV}{g} \int_{-g/2}^{g/2} \cos\omega\frac{z}{v} dz$$

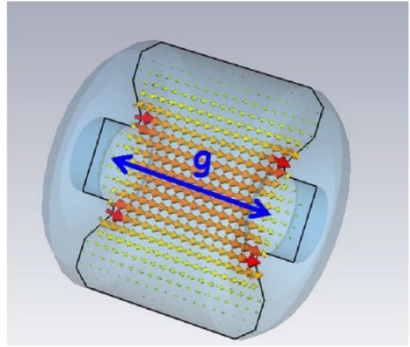

**Fig. 9**: Typical RF resonator with calculated field vector and gap size g

As usual $z$ refers here to the longitudinal direction, $V$ is the maximum amplitude of the applied voltage and $v$ is the speed of the particle. The integration refers to the complete distance between the cavity walls (the *gap*), and we refer for symmetry reasons to the centre of the RF resonator. Solving the integral we obtain

$$\Delta W = eV \frac{\sin\theta/2}{\theta/2} = eVT$$

The parameter $\theta$ is called transit angle and is defined as

$$\theta = \frac{\omega g}{v}$$

and the transit time factor $T$, describing the effective available accelerating voltage in an RF system, is

$$T = \frac{\sin\theta/2}{\theta/2}$$

The ideal case of a transit time factor approaching 1 is obtained if the argument $\theta/2$ is as small as possible, ideally $\theta/2 \rightarrow 0$, which corresponds either to $\omega \rightarrow 0$ as in the case of d.c. acceleration or to $g \rightarrow 0$. This means that we obtain the most effective use of the RF system for the smallest gap distance of the cavity. The simple case shown in Fig. 8 therefore is not an ideal situation. An improved cavity design will rather look like the one in Fig. 10, where the gap is reduced to the limit of discharges that will occur if the resulting electric field is pushed too high.

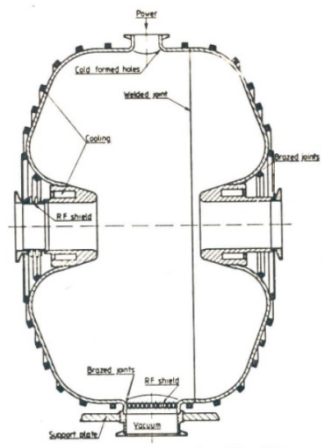

**Fig. 10:** Example of an optimised cavity design to increase the transit time factor

# 6 Synchrotrons

A synchrotron is a circular accelerator or storage ring with

- a design orbit of constant radius defined by the arrangement and strength of a number of dipole magnets that according to Eq. (2) define the particle rigidity (or momentum),
- an RF system, located at a distinct place in the ring and powered at an RF frequency that is equal to the revolution frequency of the particles or an integer multiple (so-called harmonic) of it.

For the description of the particle dynamics, we refer to a synchronous particle of ideal energy, phase and energy gain per turn. As we will see, the synchronization between the RF system and the particle beam is of major importance in this machine and has to be explicitly included in the design. To understand its principle, we have to refer briefly to the transverse dynamics of a particle with momentum error and the resulting dispersive effects (Fig. 11).

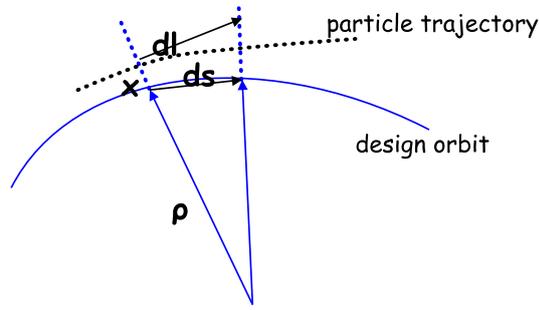

**Fig. 11:** Trajectories for particles of design energy and off momentum

While the ideal particle will run on the design orbit defined by the dipole magnets and will proceed a distance d$s$, a non-ideal particle running on a displaced orbit (displaced to the outer side of the ring in the example of Fig. 11) will pass the corresponding distance d$l$:

$$\frac{dl}{ds} = \frac{\rho + x}{\rho}$$

Solving for d$l$ we obtain

$$dl = \left(1 + \frac{x}{\rho(s)}\right) ds$$

and integrating around the machine we get the orbit length of the non-ideal particle, which depends on the radial displacement $x$:

$$l_{\Delta E} = \int dl = \int \left(1 + \frac{x_{\Delta E}}{\rho(s)}\right) ds$$

where we assume that the radial displacement $x_{\Delta E}$ is caused by a momentum error and the dispersion function of the magnet lattice

$$x_{\Delta E}(s) = D(s)\frac{\Delta p}{p} \qquad (4)$$

we obtain an expression for the difference in orbit length between the ideal and dispersive particle, which is determined by the amount of relative momentum error and the dispersion function in the storage ring:

$$\delta l_{\Delta E} = \frac{\Delta p}{p} \int \left(\frac{D(s)}{\rho(s)}\right) ds$$

The ratio between relative orbit difference and relative momentum error is called the momentum compaction factor $\alpha_p$ and using Eq. (4) it is determined by the integral of the dispersion function around the ring and the bending radius of the dipole magnets:

$$\frac{\delta l_\varepsilon}{L} = \alpha_p \frac{\Delta p}{p}$$

$$\alpha_p = \frac{1}{L} \oint \left(\frac{D(s)}{\rho(s)}\right) ds$$

For first estimates, let us assume equal bending radii in all dipoles, so $1/\rho$ = const and we replace the integral of the dispersion around the ring by a sum over the average dispersion in the dipole magnets (outside the dipoles the term $1/\rho = 0$, so this assumption is justified for a rough estimate):

$$\int_{dipoles} D(s)\, ds \approx l_{\Sigma(dipoles)} \cdot \langle D \rangle_{dipole}$$

So we get a nice and simple expression for the momentum compaction factor that depends only on the ratio of average dispersion and geometric radius $R$ of the machine:

$$\alpha_p = \frac{1}{L} l_{\Sigma(dipoles)} \cdot \langle D \rangle \frac{1}{\rho} = \frac{1}{L} 2\pi\rho \cdot \langle D \rangle \frac{1}{\rho}$$

$$\alpha_p \approx \frac{2\pi}{L} \langle D \rangle \approx \frac{\langle D \rangle}{R}$$

Assuming finally that the particles are running at the speed of light, $v \approx c$ = const, the ratio between relative error in time is given by the relative change of the orbit length and thus by the momentum compaction factor and the relative momentum error:

$$\frac{\delta T}{T} = \frac{\delta l_\varepsilon}{L} = \alpha_p \frac{\Delta p}{p} \tag{5}$$

## 6.1 Dispersive effects in synchrotrons

In the case of non-relativistic particles, however, and the problem of synchronization, we need a more careful treatment, which will give us a relation between the revolution frequency and the momentum error of the particles. The parameter of interest is the ratio between the relative momentum error and the relative frequency deviation of a particle:

$$\frac{df_r}{f_r} = \eta \frac{dp}{p}$$

Given the revolution frequency as a function of machine circumference and speed,

$$f_r = \frac{\beta c}{2\pi R}$$

we obtain via logarithmic derivation

$$\frac{df_r}{f_r} = \frac{d\beta}{\beta} - \frac{dR}{R} \tag{6}$$

Remembering that the relative change in radius, i.e., the second term in the expression, is given by the momentum compaction factor $\alpha_p$,

$$\frac{dR}{R} = \alpha \frac{dp}{p}$$

The momentum is related to the particle energy and its velocity

$$p = mv = \beta\gamma \frac{E_0}{c}$$

and its relative error is obtained as

$$\frac{dp}{p} = \frac{d\beta}{\beta} + \frac{d(1-\beta^2)^{-\frac{1}{2}}}{(1-\beta^2)^{-\frac{1}{2}}} = (1-\beta^2)^{-1} \frac{d\beta}{\beta}$$

Introducing the last two equations into Eq. (6), we finally obtain the required relation between frequency offset and momentum error

$$\frac{df_r}{f_r} = \left(\frac{1}{\gamma^2} - \alpha\right)\frac{dp}{p} \qquad (7)$$

which for ultra-relativistic particles reduces to the simplified expression of Eq. (5). Accordingly the $\eta$ parameter defined above is given as

$$\eta = \frac{1}{\gamma^2} - \alpha$$

As an important remark we state that the change of revolution frequency depends on the particle energy $\gamma$ and possibly changes sign during acceleration. Particles get faster in the beginning and arrive earlier at the cavity location (classical regime), while particles that travel at $v \approx c$ will not get faster any more but rather get more massive and, being pushed to a dispersive orbit, will arrive later at the cavity (relativistic regime). The boundary between the two regimes is defined for the case where no frequency dependence on d$p/p$ is obtained, namely, $\eta = 0$ and the corresponding energy is called the 'transition energy':

$$\gamma_{tr} = \frac{1}{\sqrt{\alpha}}$$

In general we will design machines in such a way as to avoid the crossing of this transition energy. As it involves changes of the RF phase unless, the particles lose their longitudinal focusing created by the sinusoidal RF function, the bunch profile will dilute and get lost. Qualitatively the longitudinal focusing effect and the problem of gamma transition is explained in Fig. 12.

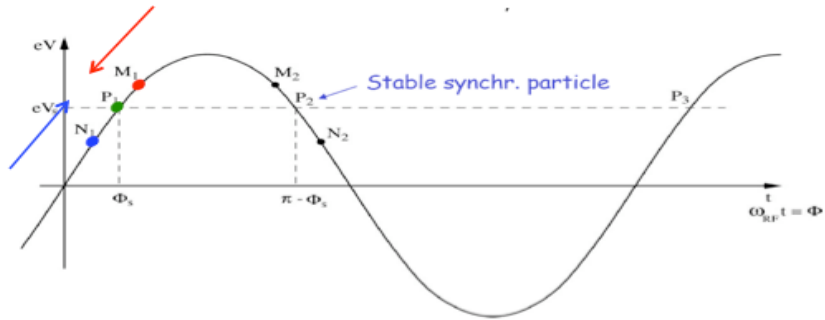

**Fig. 12:** Qualitative picture of the phase focusing principle below transition

## 6.2 The classical regime

Assume an ideal particle that is passing the cavity in time (or phase) the resonator at a certain position, as indicated by the green spot in Fig. 12. It will see a certain accelerating voltage and correspondingly receive an energy increase. We call this phase the synchronous phase, as indicated in the plot. A particle that has smaller energy than the ideal one will travel at lower speed and arrive after the next turn later, or at a larger phase and thus see a stronger accelerating voltage. It will therefore compensate the lack in energy and step-by-step come closer to the ideal particle. Just the opposite happens to a particle that has a positive energy offset. As it is faster than the synchronous particle, it will arrive earlier at the cavity, see a smaller voltage and again step-by-step will approach the ideal particle. In both cases a net focusing effect is obtained, which is due to the relation between momentum and speed and the right choice of the synchronous phase.

For highly relativistic particles the same effect exists but based now on the mass increase with energy. As visualized in Fig. 13 the high-energy particle (marked in blue) will, due to its higher mass, run on a longer orbit and – as its speed is constant $v \approx c$ – it will arrive later at the cavity location.

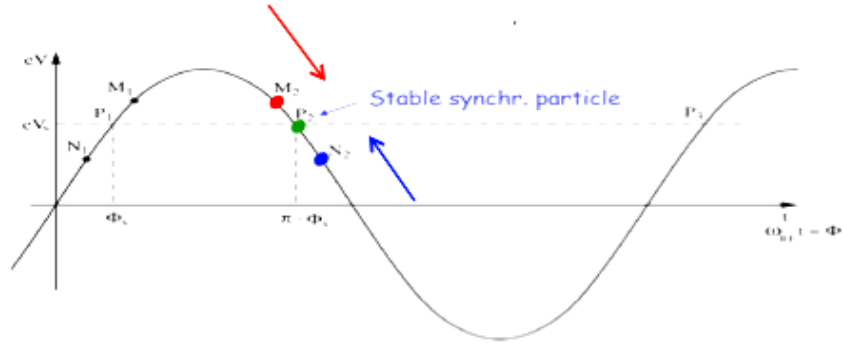

**Fig. 13:** Phase focusing situation above transition

As a consequence, the synchronous phase has to be chosen depending on whether we are running the machine below or above transition. Synchrotrons that have to pass through transition will have to apply a phase jump to keep the particles bunched.

In this context it is worth having a look at the acceleration mechanism itself. The particle momentum is defined via the beam rigidity by the dipole field $B$,

$$p = eB\rho$$

and as a consequence a change in particle momentum is reflected by an appropriate change of the $B$ field:

$$\frac{dp}{dt} = e\rho \dot{B}$$

The momentum increase per turn is therefore given by

$$(\Delta p)_{turn} = e\rho \dot{B} T_r = \frac{2\pi e\rho R \dot{B}}{v}$$

and referring to the energy change rather than the change in momentum we get with

$$E^2 = E_0^2 + p^2 c^2 \quad \Rightarrow \quad \Delta E = v\Delta p$$

the change in energy per turn, which is clearly related to the accelerating voltage and the synchronous phase $\phi_s$ of the particles:

$$\Delta E_{turn} = \Delta W_{turn} = 2\pi e\rho R\dot{B} = e\hat{V}\sin\phi_s$$

The following are some remarks that might be worth emphasizing:
- The energy gain depends on the rate of change of the dipole field.
- The number of stable synchronous particles is equal to the harmonic number $h$. They are equally spaced along the circumference.
- Each synchronous particle satisfies the relation $p = eB\rho$. They have the nominal energy and follow the nominal trajectory.
- As long as the particles are not fully relativistic yet, their revolution frequency will change and so does the RF frequency, which is a multiple of $f_r$ to stay in synchronization during the complete acceleration process.

### 6.3 Frequency change during acceleration

The relation between the revolution frequency and RF frequency is defined by the harmonic number and depends on the size of the ring and the magnetic dipole field:

$$f_r = \frac{f_{RF}}{h} = f(B, R_s)$$

Hence, using the beam rigidity relation and the average dipole field to define the radius of the ideal particle, we get for an average magnetic field $<B(t)>$

$$\frac{f_{RF}(t)}{h} = \frac{v(t)}{2\pi R_S} = \frac{1}{2\pi}\frac{e}{m} <B(t)>$$

$$\frac{f_{RF}(t)}{h} = \frac{1}{2\pi}\frac{ec^2}{E_S(t)}\frac{r}{R_S} B(t)$$

Using the relativistic overall energy

$$E^2 = (m_0 c^2)^2 + p^2 c^2$$

we get finally an expression for the RF frequency as a function of the changing external dipole field:

$$\frac{f_{RF}(t)}{h} = \frac{c}{2\pi R_S}\left\{\frac{B(t)^2}{(m_0 c^2/ecr)^2 + B(t)^2}\right\}^{1/2}$$

At high energies, or, more accurately, as soon as

$$B > \frac{m_0 c^2}{ecr} \tag{8}$$

the situation simplifies a lot and the equation above reduces to

$$\frac{f_{RF}(t)}{h} \approx \frac{c}{2\pi R_S} = const \tag{9}$$

so that we can keep the RF frequency constant. It is evident from Eq. (8) that for electron beams in general the condition (9) is (nearly) always fulfilled while proton or heavy-ion synchrotrons need a more sophisticated RF control.

## 7 Synchrotron motion

In the following we will again go through the longitudinal motion. However, unlike the previous section, we will try to put things on a mathematically more solid ground.

As in Fig. 12 we expect a longitudinal oscillation in phase and energy under the influence of the focusing mechanism explained above. The relation between relative frequency deviation and relative momentum error is given by Eq. (7) as

$$\frac{df_r}{f_r} = \left(\frac{1}{\gamma^2} - \alpha\right)\frac{dp}{p}$$

which translates into the difference in revolution time

$$\frac{dT}{T_0} = \left(\alpha - \frac{1}{\gamma^2}\right)\frac{dp}{p}$$

and leads to a difference in phase at the arrival at the cavity:

$$\Delta\psi = 2\pi\frac{\Delta T}{T_{rf}} = \omega_{rf} * \Delta T$$
$$= h * \omega_0 * \Delta T = h * 2\pi\frac{\Delta T}{T_0}$$
$$= h * 2\pi\left(\alpha - \frac{1}{\gamma^2}\right)\frac{dp}{p}$$
$$= \frac{h * 2\pi}{\beta^2}\left(\alpha - \frac{1}{\gamma^2}\right)\frac{dE}{E}$$

As before, the revolution frequency $\omega_0$ and the RF frequency $\omega_{RF}$ are related to each other via the harmonic number $h$. Hence the difference in energy and the offset in phase are connected to each other through the momentum compaction factor, or better the $\eta$ parameter.

Differentiating with respect to time gives the rate of change of the phase offset per turn:

$$\Delta\dot\psi = \frac{\Delta\psi}{T_0} = \frac{h * 2\pi}{\beta^2 T_0}\left(\alpha - \frac{1}{\gamma^2}\right)\frac{dE}{E} \tag{10}$$

On the other hand, the difference in energy gain of an arbitrary particle that has a phase distance of $\Delta\psi$ to the ideal particle is given by the voltage and phase of the RF system:

$$\Delta E = e * U_0(\sin(\psi_s + \Delta\psi) - \sin\psi_s)$$

As before, we describe the phase of the ideal ('synchronous') particle by $\psi_s$ and the phase difference by $\Delta\psi$. Referring to small amplitudes $\Delta\psi$ of the phase oscillations, we can simplify the treatment by assuming

$$\sin(\psi_s + \Delta\psi) - \sin\psi_s = \sin\psi_s \underbrace{\cos\Delta\psi}_{\approx 1} - \cos\psi_s \underbrace{\sin\Delta\psi}_{\Delta\psi} - \sin\psi_s$$

and get for the rate of energy change per turn:

$$\Delta\dot E = e * \frac{U_0}{T_0}\Delta\psi \cos\psi_s$$

A second differentiation with respect to time delivers

$$\Delta\ddot{E} = e * \frac{U_0}{T_0} \Delta\dot{\psi} \cos\psi_s \qquad (11)$$

Combining Eqs. (10) and (11) we get finally a differential equation for the longitudinal motion under the influence of the phase focusing mechanism:

$$\Delta\ddot{E} = e * \frac{U_0}{T_0} \frac{2\pi h}{\beta^2 T_0} \left(\alpha - \frac{1}{\gamma^2}\right) \frac{dE}{E} \cos\psi_s$$

For a given energy the parameters in front of the right-hand side are constant and describe the longitudinal – or 'synchrotron' – oscillation frequency. Using therefore

$$\Omega = \omega_0 * \sqrt{\frac{-eU_0 h \cos\psi_s}{2\pi\beta^2 E}\left(\alpha - \frac{1}{\gamma^2}\right)} \qquad (12)$$

we get the equation of motion in the approximation of small amplitudes:

$$\Delta\ddot{E} + \Omega^2 \Delta E = 0$$

It describes a harmonic oscillation in $E$–$\psi$ phase space for the difference in energy of a particle to the ideal (i.e., synchronous) particle under the influence of the phase focusing effect of our sinusoidal RF function.

As already discussed qualitatively, the expression (12) leads to real solutions if the argument of the square root is a positive number. Two possible situations therefore have to be considered: below the gamma transition the $\eta$ parameter is positive, and above it $\eta$ is negative. The synchronous phase as the argument of the cos function therefore has to be chosen to get an overall positive value under the square root:

$$\gamma < \gamma_{tr} \qquad \eta > 0, \quad 0 < \phi_s < \pi/2$$
$$\gamma > \gamma_{tr} \qquad \eta < 0, \quad \pi/2 < \phi_s < \pi$$

As an example Fig. 14 shows the superconducting RF system of the LHC, and the basic parameters, including the synchrotron frequency, are listed in Table 1.

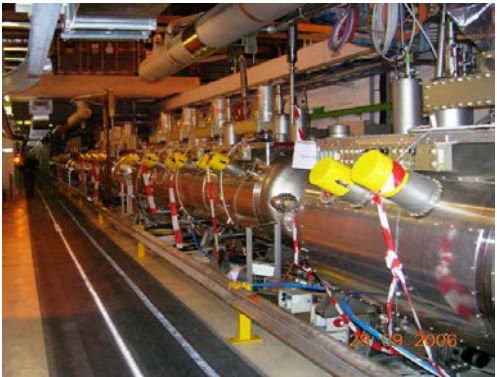

**Fig. 14:** LHC RF system

**Table 1:** Parameters of the LHC RF system.

| | | |
|---|---|---|
| Bunch length (4σ) | ns | 1.06 |
| Energy spread (2σ) | $10^{-3}$ | 0.22 |
| Synchrotron rad. loss / turn | keV | 7 |
| RF frequency | MHz | 400 |
| Harmonic number $h$ | | 35640 |
| RF voltage / beam | MV | 16 |
| Energy gain / turn | keV | 485 |
| Synchrotron frequency | Hz | 23 |

Figure 15 finally shows the longitudinal focusing effect that has been observed during the LHC commissioning phase. On the left-hand side, beam had been injected into the storage ring while the RF system was still switched off. The bunch, nicely formed by the RF voltage of the pre-accelerator, is visible only for a few turns and the bunch profile is decaying fast, as no longitudinal focusing is active. The right-hand side shows the situation with the RF system activated and the phase adjusted. The injected particles stay nicely bunched and the acceleration process can start.

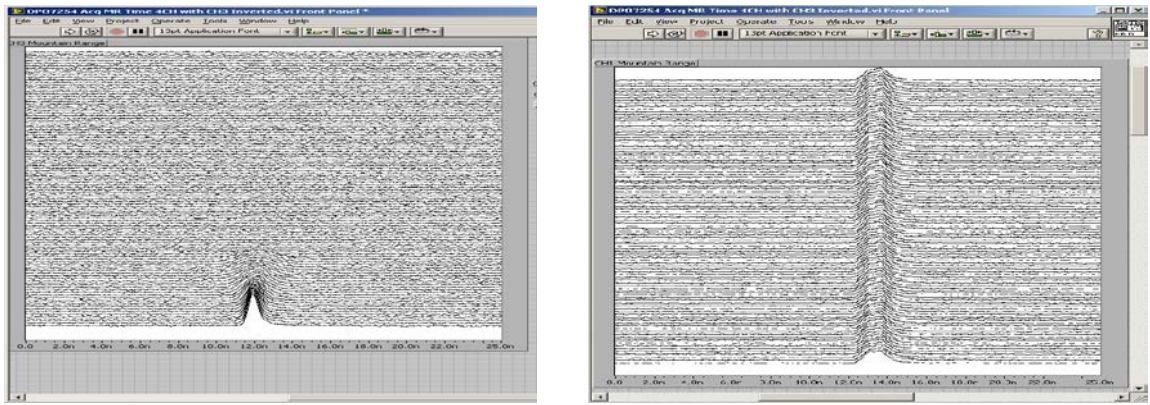

**Fig. 15**: Longitudinal beam profile measured during LHC injection. Left: the RF system is switched off, and the bunch profile is decaying fast. Right: the RF system is activated, and the particles stay bunched.